\documentclass[twocolumn,showpacs,preprintnumbers,amsmath,amssymb]{revtex4}

\usepackage{graphicx}
\usepackage{dcolumn}
\usepackage{bm}

\newcommand{\be}{\begin{equation}}
\newcommand{\ee}{\end{equation}}

\begin{document}

\preprint{CGG-08-07-19}

\title{On the shoulders of Giants - Quantum gravity and braneworld stability}

\author{Alex Hamilton}
\author{Jeff Murugan}
\affiliation{Cosmology \& Gravity Group, Department of Mathematics and Applied Mathematics, Univeristy of Cape Town,Private Bag, Rondebosch 7700, South Africa}


\begin{abstract}
\noindent
The semi-classical nature of braneworld cosmological models does not account for any quantum gravitational effects. In this letter we use the gauge/gravity correspondence to argue that quantum string corrections {\it cannot} be ignored in any study of braneworld stability. As an example, we find, by analysing the quantum gravitational backreaction, that a closed universe is unstable to radiation into the bulk.
\end{abstract}


\maketitle

The idea of a gauge/gravity duality - the equivalence of two markedly different systems containing different fields, degrees of freedom and even a differing number of spacetime dimensions - has revolutionized string theory. Indeed, in the ten years since the discovery of its most concrete form, the AdS/CFT conjecture \cite{Maldacena}, it has illuminated such notoriously difficult problems as black hole radiation, gravitational entropy, strongly coupled QCD physics and even the emergence of locality \cite{Developments}. 
In this light it is quite remarkable that, barring a few outstanding examples \cite{Examples}, this extraordinary tool has not made more of an impact in cosmology. 
In contrast, braneworld models, largely unfettered by the rigid constraints imposed by string theory, have generated enormous interest among both particle physicists and cosmologists alike \cite{Braneworlds}. In this letter, we attempt to redress some of this inequity by (a) showcasing AdS/CFT as a possible tool to study cosmological questions and (b) applying this correspondence in a controlled setting to study essentially quantum gravitational corrections to braneworld dynamics. 

Specifically, we work in a 10-dimensional $AdS_{5}\times S^{5}$ string background, and study the `mirage' cosmology \cite{Mirage} induced on a giant graviton (a spherical D3-brane extended in the S${}^{3}$ of AdS${}_5$) which moves within the transverse five-sphere. An observer on the D-brane worldvolume sees an Einstein static universe whose size is determined by the angular momentum of the brane \cite{Youm}. We will show that, unlike the Einstein static solution in general relativity \cite{Eddington}, this worldvolume cosmology is semiclassically neutrally stable to homogeneous perturbations - a desirable property in several recent contexts \cite{Emergent}. However - and this is the key element of our analysis - a careful accounting of the stringy degrees of freedom in the problem reveals an  instability of the braneworld in the form of gravitational radiation into the bulk. It is important to note (a) that this is a quantum effect made tractable through the AdS/CFT correspondence and (b) our argument is quite generic. The D-brane is a charged accelerating object whose gravitational radiation cannot be ignored. We expect that this effect could have important repercussions for other braneworld systems based on D-branes \cite{KKLMMT}.\\
 
To reiterate, the background on which the giant graviton moves is AdS${}_5 \times S^5$ with metric
$
ds^2_{\mathrm{bulk}} = -\left( 1+\frac{r^2}{R^2} \right) dt^2 + \left(1+\frac{r^2}{R^2} \right)^{-1} dr^{2} +r^2 d\Omega_3^2 + R^2 d\tilde{\Omega}_5^2,
$
sourced by the energy density of an accompanying ``magnetic'' four-form field $C_{(4)}$ (the Ramond-Ramond flux), to which the D3-brane couples.  In this setup, we consider a spherical D3-brane which is embedded in the 3-sphere of the AdS space at some (in general, time dependent) radius $r$. Naively, the tension of the brane would cause an immediate collapse. However, the fact that this string-theoretic object is charged under the $C_{(4)}$ field allows it to expand when moving - the higher dimensional analogue of a dipole moving in a magnetic field \cite{Giants}. The dynamics of the D-brane in this background is encoded in its action
\be
  S =
   -T_{3}\!\!\!\!\!\!\!\underbrace{\int\,d^{4}\xi\,\sqrt{|\hat{g}|}}_{\mathrm{Dirac-Born-Infeld}}\,\,\, +\,\,\, T_{3}\!\!\!\!\!\underbrace{\int\,C_{(4)}}_{\mathrm{Chern-Simons}} \,,
  \label{DBI action}
\ee
where $T_{3}$ is the tension/charge of the brane and $\hat{g}$ is the induced metric.  Setting the time coordinate along the brane $\xi^0 = t$ (the static gauge choice) and assuming that the brane has (conserved) angular momentum $l$ along the $\varphi$ direction of the $S^5$, gives the energy
\begin{eqnarray}
E = \left( 1+ \frac{r^2}{R^2} \right)^{3/2} \sqrt{\frac{(\rho r^3)^2 + (l/R)^2}{(1+\frac{r^2}{R^2})^2 - \dot{r}^2}} - \frac{\rho r^4}{R}\,,
\end{eqnarray}
of the D-brane as well as the time evolution of its radius
\be
\label{Radial-eqn}
\dot{r}^2 = \left( 1 + \frac{r^2}{R^2} \right)^2 \left[ 1 - \left( 1 + \frac{r^2}{R^2} \right) \frac{(\rho r^3)^2 + (l/R)^2}{\left( E + \frac{\rho r^4}{R} \right)^2} \right]\,,
\ee
and five-sphere coordinate
\be
\label{Angular-eqn}
\dot{\varphi}^{2} = \frac{l^2}{R^{4}} \left( 1 + \frac{r^2}{R^2} \right)^{2}\left( E + \frac{\rho r^4}{R} \right)^{-2} \,.
\ee
Here, $\rho = 2\pi^2 T_3$ can be thought of as a mass density for the sphere (i.e., $m = \rho r^3$).

To say anything cosmological about the brane, we need to first translate to the worldvolume point of view. The induced metric in an embedding $X^A$ may be found from $\hat{g}_{\mu \nu} = g_{AB} \ \partial_\mu X^A \partial_\nu X^B$ using (\ref{Radial-eqn}) and (\ref{Angular-eqn})  
\begin{eqnarray}
  ds^{2} 
  = -\left[\frac{(\rho r^{3})^2 \left(1+ r^{2}/R^{2}\right)^{2}}{\left(E+ \rho r^{4}/R\right)^{2}}\right]\,
  dt^{2} + r^{2}d\Omega_{3}^{2}\,.
\end{eqnarray}
Following \cite{Youm}, the first term in this expression can be used to define the `cosmic' time measured by worldvolume observers with an induced metric, $ds^{2} = -d\tau^{2} + r^{2}\,d\Omega_{3}^{2}$ for a closed ($k=+1$) FRW-like universe with scale factor $a(\tau)\equiv r(\tau)$. Now, using the expression for $\dot{r}$, denoting derivatives with respect to the cosmic time $\tau$ by a prime and 
rescaling the energy and angular momentum by a constant for convenience, gives the effective Friedmann equations for the scale factor
\begin{eqnarray}
  \left(\frac{a'}{a}\right)^{2} &=& \frac{1}{R^{2}}\left[\left(E^{2} - 
  l^{2}\right)X^{-8} + 2EX^{-4}
  - X^{-2}
  - l X^{-6}\right],\nonumber\\
  \frac{a''}{a} &=& -\frac{1}{R^{2}}\left[3\left(E^{2} - 
  l^{2}\right)X^{-8} + 2EX^{-4} - 2l 
  X^{-6}\right] \;, \nonumber
\end{eqnarray}
where we have defined the dimensionless $X\equiv a/R$. Writing the evolution equations in this way, allows for the right-hand side to be interpreted as `mirage' matter with effective energy density and pressure
\begin{eqnarray}
  \varrho_{\rm eff} &=& \frac{3}{8\pi R^{2}}\left[\left(E^{2} - 
  l^{2}\right)X^{-8} + 2EX^{-4} 
  - l X^{-6}\right],\nonumber\\
  p_{\rm eff} &=& \frac{1}{8\pi R^{2}}\left[ 5\left(E^{2} -
  l^{2}\right)X^{-8} + 2EX^{-4} 
  - 3l X^{-6}\right],\nonumber
\end{eqnarray}
respectively. The effective energy density of the induced matter on the D-brane worldvolume recieves three separate contributions depending on how it scales with the cosmic scale factor $a$.  There is radiation ($\varrho \sim a^{-4}, w=1/3$) as well as a massless scalar term ($\varrho \sim a^{-6}, w=1$).  Finally, the acausal term ($\varrho \sim a^{-8}, w=5/3$) is a telltale indicator that the matter cannot be real - there must be some sort of mirage cosmology happening.\\

We now look for static solutions to the worldvolume cosmology. Such solutions necessarily satisfy $a' = a'' = 0$. From the first Friedman equation, $H^{2}=0$ if,
\begin{eqnarray}
  X^{3} -2EX^{2} +l^{2}X - (E^{2}-l^{2})=0.
  \label{Dimensionless-F1}
\end{eqnarray}
It is not hard to see \cite{Giants} that the lowest energy state of this system is an extremal (or BPS) configuration with $E=l$ and two possible solutions for the size. Not surprisingly, the two BPS solutions that satisfy (\ref{Dimensionless-F1}) with $E=l$ are $X=0$ and $X=l$. The first of these, corresponding to system of vanishing size, is known to be a KK-graviton state in the string theory, following a geodesic around a great circle of the $S^5$. The latter is the {\it giant graviton} blown up along the 3-sphere in $AdS_{5}$, with scale factor $a_{0}=R\sqrt{l}$ in our normalisation. As a check that this really does correspond to a static solution, note that this solution leads to $H'=0$ from the second effective Friedmann equation. From the worldvolume point of view then, the AdS giant graviton is an Einstein static-like universe with $k=+1$, $a'=a''=0$ but with {\it no cosmological constant} term usually required to counterbalance the gravitational collapse of the Universe. Here, the D3-worldvolume tension is supported by the angular momentum of the giant on the $S^{5}$ through a coupling to the RR 4-form flux. We demonstrate now that, unlike in General Relativity, this static universe does not suffer the same instability to homogeneous perturbations.\\

Any homogeneous (and isotropic) perturbation around the $a_{0}=R\sqrt{l}$ fixed point is of the form 
$a(\tau) = a_{0} + \epsilon\alpha(\tau)$. To check the stability of the solution we follow \cite{Eddington}, expand all quantities in the effective Hubble evolution equation $ 
 H' = -4\pi(\varrho_{\rm eff} + p_{\rm eff}) + \frac{1}{a^{2}}
$ in $\epsilon$ and look for a growth in the fluctuation. 
To linear order, this prescription leads to an undamped oscillator equation for the fluctuation
\begin{eqnarray}
  \alpha''(\tau) = -\frac{4}{R^{2}l}\alpha(\tau).
  \label{perturbation}
\end{eqnarray}
Evidently, the AdS giant cosmology, unlike the usual Einstein static solution of general relativity, is {\it  stable} to homogeneous perturbations.  However, without an exit channel the energy of the fluctuation, an arbitrary perturbation will remain undamped. To check that this isn't merely an artifact of the linearization, we have also numerically integrated the dynamical system defined by the Hubble evolution and Energy conservation equations with initial conditions subject to the Friedmann constraint. The results, plotted in FIG.1, display excellent agreement with the linearization analysis.\\

\begin{figure}[h] 
   \centering
   \includegraphics[width=5cm]{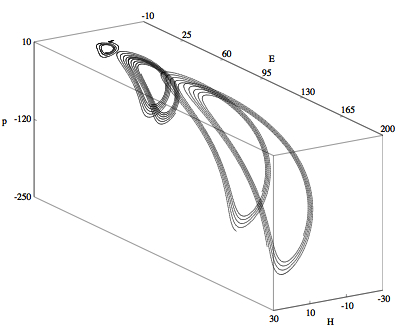} 
   \caption{Numerical integration of the Friedman equations show that the giant worldvolume is a stable center of the ($\rho=E$,$p$,$H$) phase space.} 
   \label{DS}
\end{figure}

\noindent
Such marginal solutions are not unique to string theory and have been shown to exist everywhere from general relativity through to loop quantum cosmology and various modified gravity models \cite{New_Einstein_Static}. However, the story is a little more subtle. Thus far we have only analysed the spectrum of small fluctuations about the D3-brane worldvolume at the level of the DBI action (\ref{DBI action}). Such fluctuations are open strings attached to the D-brane. The DBI action, however, encodes just the low energy, semi-classical degrees of freedom, corresponding to the zero mode sector of the string theory. 
String interactions bring qualitative differences in brane behavior. In particular, it was observed in \cite{Berenstein} and verified in spectacular detail in the series of articles \cite{deMelloKoch} that among several remarkable phenomena is a runaway growth of open strings attached to giant gravitons. Qualitatively, this arises because the giant graviton, as a $C_{(4)}$-charged soliton in the theory, couples to the 4-form flux and is hence accelerated relative to the background. Strings, on the other hand, are uncharged, and so should be in geodesic freefall. Since they are attached however, they are dragged along the accelerated path of the giant. In response, these open strings experience a force that both stretches them and brings their endpoints closer together. When these endpoints meet, the resulting {\it closed string}, no longer subject to the Dirichlet boundary conditions that fix it to the D-brane, will separate from the giant and radiate into the AdS geometry. Since, in string theory closed strings encode gravitational degrees of freedom, this process of strings peeling off the D-brane is most naturally interpreted as {\it gravitational radiation} from the braneworld into the bulk.  

To make the argument more quantitative, we now turn to quantum gravity effects.  Key to our computation is the exact AdS/CFT correspondence in string theory, by which such quantum gravitational effects are encoded in a more tractable way in a dual field theory. 
Herein lies the main feature that distinguishes the giant graviton from other braneworld models: not only are the states dual to the D-brane known precisely, but the correlators that measure the way they propagate and interact can be computed exactly in the large $N$ limit \cite{Corley}. Given that the correspondence predicts a string coupling constant $g_s \sim 1/N$, we now have the tools to understand first order quantum gravitational effects - in particular, the decay of brane fluctuations. 
From string theory perspective, homogeneous fluctuations are equivalent to open string excitations oscillating transversely to the brane 
\footnote{Classically, such individual strings would not give rise to homogeneous fluctuations.  Once quantized, however, it is precisely the low energy s-wave excitations which are homogeneous, and it is with these that we are concerned.}.  
To understand the decay of these fluctuations, we must therefore first identify the relevant open string states, as well as their decay products.
In particular, we are interested in the low energy excitations. The obvious decay candidates are bulk gravitons, with a spectrum of KK modes deriving from the compactified five-sphere.  Parameterizing the $S^{5}$ by three complex coordinates, $X,Y,Z$ such that $|X|^2 + |Y|^2 + |Z|^2=1$, such modes are characterized by three angular momenta around the sphere, $(J_X, J_Y, J_Z)$ where $J_X$ corresponds to angular momentum around the circle $|X|^2 = \mathrm{constant}$, etc.  The low energy open strings will likewise oscillate along the directions of the five-sphere with the same momentum ``charges.''

These, then, are the elements of the interaction on which we will focus - the initial state, $| \mathrm{i} \rangle$, being a giant graviton with open string excitation polarized along the five-sphere, and the final state, $| \mathrm{f} \rangle$, being the giant ground state plus KK graviton.  Recalling Fermi's Golden Rule, that the decay rate is (up to phase factors) the probability of the transition
$
\Gamma \sim | \langle \mathrm{f} | \mathrm{i} \rangle |^2
$, our goal is now to find the gauge theory states corresponding to  $| \mathrm{i} \rangle$ and  $| \mathrm{f} \rangle$ and use knowledge of the dual theory to calculate their overlap \cite{Brown}.

Before identifying the dual field theory states, let us take a moment to examine the polarization directions.  Being extended solely in the AdS directions, the D3-brane exists as a point in the five-sphere, traveling around the great circle with angular momentum $l$.  We call this the $Z$ direction, and  set $J_Z = l$.  There remains, then, an SO(4) symmetry, for which we need only consider fluctuations in, say, the $Y$ direction.  This we deal with below.  However, we must also consider fluctuations along the plane of motion of the brane, in the $Z/\bar{Z}$ directions.  Results in \cite{deMelloKoch} indicate a potential instability in these fluctuations, which certainly merits further investigation.  This is a more technically formidable problem, which we deal with in a separate article \cite{Hamilton}.

States in a CFT are dual to operators, which are made up of combinations of matrices (of the group SU($N$)) from the gauge theory. In particular, there are three complex scalars, $X,Y,Z$, which correspond exactly to the directions on the five-sphere.  String states can be built from these scalars by considering each field as a string `bit' oscillating in the corresponding direction of the sphere.  The length of the string is then proportional to the number of multiplied fields, and its angular momentum on the $S^5$ related to the number of fields of each type.  Additionally, after matrix multiplication, there remain two uncontracted indices - corresponding to two ends of an open string.  In order to form a closed string, we self-contract these indices, which is equivalent to taking the trace of the string operator, or graviton.  For concreteness, we will consider a string with $J$ units of angular momentum along the $Y$ direction, $\left( Y^J \right)^i_j$.

The D3-brane is a little more complicated.
We are interested in a spherical D3-brane extended in the $S^3$ of AdS${}_5$.  The radius of the giant is quantized, $r=(p/N)R$, where $p$ is an integer and $R$ is the AdS length scale.  We have seen that in order to support this radius, the brane must have $p$ units of angular momentum on along the $Z$ direction of the five-sphere. Remarkably, it was shown in \cite{Corley} that the relevant gauge theory operator is a particular combination of $p$ of the $Z^i_j$ matrices, called a Schur polynomial, $\chi_{(p)} (Z)$.
Adding a fluctuation to this state is equivalent to attaching an open string to the D3-brane.  
Now, the relevant operator made from $p$ of the $Z$'s, and the open string word $\left( Y^J \right)^i_j$ is known as a \emph{restricted} Schur polynomial, $\chi^{(1)}_{(p+1),(p)} (Z, Y^J)$ (see \cite{deMelloKoch} and references therin).
To calculate the decay rate of an excited giant graviton state then, we start with a normalized initial state represented by
\begin{eqnarray}
| \mathrm{i} \rangle = \frac{1}{\mathcal{N}_i} \left| \chi_{(p+1),(p)}^{(1)} (Z, Y^J) \right\rangle
\end{eqnarray}
and end with a final state of a graviton plus non-excited giant\footnote{We could also consider processes in which only part of the excitation energy is radiated away as a graviton, but this process is suppressed in the large $N$ limit.}
\begin{eqnarray}
| \mathrm{f} \rangle = \frac{1}{\mathcal{N}_f} \left| \chi_{(p)} (Z) \ \mathrm{Tr} \ Y^J \right\rangle \;.
\end{eqnarray}
Following the prescription in \cite{deMelloKoch}, the overlap between them is easily computed. Up to $\mathcal{O}(1/N)$ corrections we get,
\begin{eqnarray}
\mathcal{A} = \langle \mathrm{f} | \mathrm{i} \rangle &\sim& \sqrt{\frac{J(1+p/N)}{p+J + p(J-1)/N}} \left(1 + \mathcal{O}(J^4/N^2) \right) \nonumber \\
					& \sim & \sqrt{\frac{J(1+p/N)}{p}} \left(1 + \mathcal{O}(J/N) \right) \;,
\end{eqnarray}
where in the second line it is assumed that the giant is large enough to be distinguished from normal graviton ($p \sim \mathcal{O}(N)$).
The decay rate is then easily read off, in units defined by the size of the AdS space, $1/R$, as
$\Gamma \sim | \mathcal{A} |^2 \sim  \frac{J(1+p/N)}{p} \left(1 + \mathcal{O}(J/N) \right)$. 
At the level of the worldvolume cosmology, this decay process corresponds to gravitons that carry away energy from the brane matter into the bulk, giving a non-vanishing contribution to the energy conservation equation.\\ 

We close with a summary and some comments and leave a detailed study of the modified dynamical system to future work \cite{Hamilton}. {\it We have found a braneworld in string theory that is semiclassically stable but destabilizes due to quantum corrections}. For the $k=+1$ FRW case the effect we have documented is most dramatic. The brane worldvolume is a closed surface, which means it carries no monopole charge, just a dipole (plus higher order multipoles if it is not a perfect sphere \cite{Deformed}). With no conservation law protecting the brane from dissappearing completely, this universe could completely dissolve into gravitational radiation. For k=0 or -1, the brane is no longer a giant graviton so while there are quantitative changes in the decay rate, we still expect similar radiation effects \footnote{We'd like to thank R. de Mello Koch for emphasising this to us.}. There is a monopole charge for these configurations, so the universe is ultimately stable.  However, a crinkly world carries higher moments (after the monopole) but we expect these to be unstable and decay. In the k=0 case, for example, the brane would evolve toward being as flat as possible {\it i.e.} local inhomogeneities would wash out by the mechanism we discussed here. Clearly these issues need to be addressed if braneworlds are to be considered realistic models of the universe.  
Evidently, by asking the correct questions, the AdS/CFT correspondence has the potential to be every bit as useful a tool to cosmology as it has proven in, say heavy ion physics. It is to this end that we hope that this letter will, if little else, stimulate a more cosmological audience to think about the 
gauge/gravity duality.\\

We are grateful to K. Ananda, B. Bassett, C. Clarkson, G. Ellis, J. Larena, J. Leach, R. Maartens, J-P. Uzan and A. Weltman for much stimulating discussion. In addition, a special word of thanks is extended to R. de Mello Koch for encouragement and sharing with us some of the wonders of Young diagram technology. This work is supported by NRF grant TTK2006040500022.

\end{document}